\documentclass[pra,twocolumn,amsmath,amssymb,longbibliography]{revtex4-2}
\usepackage{graphicx}
\usepackage{amssymb}
\usepackage{amsmath}
\usepackage[braket, qm]{qcircuit}
\usepackage{bbold}
\usepackage{appendix}
\usepackage{xcolor}

\newcommand{\nmax}{n}
\usepackage{bbm}
\usepackage{subcaption}
\usepackage{lineno}
\usepackage{ulem}
\begin{document}
\title{Improved entanglement entropy estimates from filtered bitstring probabilities}
\author{Avi Kaufman$^{1}$}
\author{James Corona$^{1}$}
\author{Zane Ozzello$^{1}$}
\author{Blake Senseman$^{1}$}
\author{Muhammad Asaduzzaman$^{1,2}$}
\author{Yannick Meurice$^{1}$}
\affiliation{$^1$Department of Physics and Astronomy, The University of Iowa, Iowa City, IA 52242, USA }
\affiliation{$^2$North Carolina State University, Department of Physics, Raleigh, NC 27607, USA}

 \date{\today}

\begin{abstract}
Using the bitstring probabilities of ground states of bipartitioned ladders of Rydberg atoms, we calculate the mutual information, which is a lower bound on the corresponding bipartite von Neumann quantum entanglement entropy $S^{vN}_A$. We show that in many cases these lower bounds can be improved by removing the bitstrings with a probability lower than some value $p_{min}$ and renormalizing the remaining probabilities (filtering). We propose a heuristic based on the change of the conditional entropy under filtering that very effectively improves the estimate of $S^{vN}_A$. We consider various sizes, lattice spacings and bipartitions.  Our numerical investigation  suggest that the filtered mutual information obtained with samples having just a  few thousand bitstrings can provide reasonably close estimates of  $S^{vN}_A$. We briefly discuss practical implementations with QuEra's Aquila device.

 \end{abstract}

\maketitle

\def\beq{\begin{equation}}
\def\enq{\end{equation}}
\def\nq{n_q}
\def\nmax{n_{\mathrm{max}}}
\def\phix{\hat{\phi} _{\bf  x}}
\def\nq{n_q}
\def\ah{\hat{a}}
\def\ahd{\hat{a}^\dagger}
\def\pn{P_{\nmax-1}}
\def\har{\hat{H}^{\mathrm{har.}}}
\def\hanh{\hat{H}^{\mathrm{anh.}}}
\def\hpf{\hat{\phi}^4}
\def\np{\mathcal{N}_p}
\def\art{Article\ }
\section{Introduction} 
Recent progress in the controlled manipulation of small quantum systems has opened the possibility of using quantum devices to study aspects of lattice models considered in condensed matter or high energy physics which cannot be obtained efficiently with classical computing \cite{Bauer:2022hpo,Altman:2019vbv}. In this context, arrays of Rydberg atoms have been used to build quantum simulators for lattice gauge theory models \cite{Zhang:2018ufj,Surace:2019dtp,Notarnicola:2019wzb,Celi:2019lqy,Meurice:2021pvj, Fromholz:2022ymy,Gonzalez-Cuadra:2022hxt,
Heitritter:2022jik,Halimeh:2023lid,Gonzalez-Cuadra:2024xul}. A ladder of Rydberg atoms was initially proposed to simulate scalar QED \cite{Meurice:2021pvj}, even though it does not provide an exact match to the target model \cite{Zhang:2023agx}. But it has a very rich phase diagram that includes Berezinskii–Kosterlitz–Thouless transition, Conformal Field Theory critical points and incommensurate (floating) phases \cite{floating,Liao:2024edx}. This offers multiple opportunities to define continuum limits when the correlation lengths are large compared to the lattice spacing. Exploring large parameter spaces to identify the critical regions requires efficient methods. 

An important quantity used to study phase transitions and the critical behavior is the entanglement entropy \cite{amico2008entanglement,Eisert:2008ur,Ryu:2006bv,Abanin_2019,Cirac:2020obd,Ghosh:2015iwa,VanAcoleyen:2015ccp,Banuls:2017ena,Knaute:2024wfh, Kharzeev:2017qzs, PhysRevD.98.054007,Zhang:2021hra,Beane:2018oxh,Robin:2020aeh, PhysRevLett.90.227902,PhysRevLett.92.096402,Calabrese:2005zw}. 
The experimental measurement of entanglement entropy is challenging due to its 
nonlocal nature. One possibility is to prepare and interfere copies 
of the original system \cite{PhysRevLett.109.020504, PhysRevLett.109.020505,Islam:2015mom,Kaufman:2016mif,PhysRevLett.90.227902,PhysRevLett.92.096402,Calabrese:2005zw,Unmuth-Yockey:2016znu}. This is experimentally feasible, however it remains labor consuming and not the ideal tool for extensive phase space exploration. Recently, it was empirically observed \cite{Meurice:2024fya} that for 
arrays of Rydberg atoms, the classical mutual information 
$I_{AB}^X$ for a bipartition $AB$ obtained from the easily accessible bitstring 
distributions of a single copy, provides lower bounds that are rather close to $S^{vN}_A$, the quantum von Neumann entropy. 
The lower bound property 
\beq I^X_{AB}\leq S_{B}^{vN}=S_{A}^{vN},
\label{eq:bound}
\enq
follows from the Holevo bound \cite{mehdi,Meurice:2024fya,Yang:2024yxu}.
\begin{figure}[h]
\includegraphics[width=8.6cm]{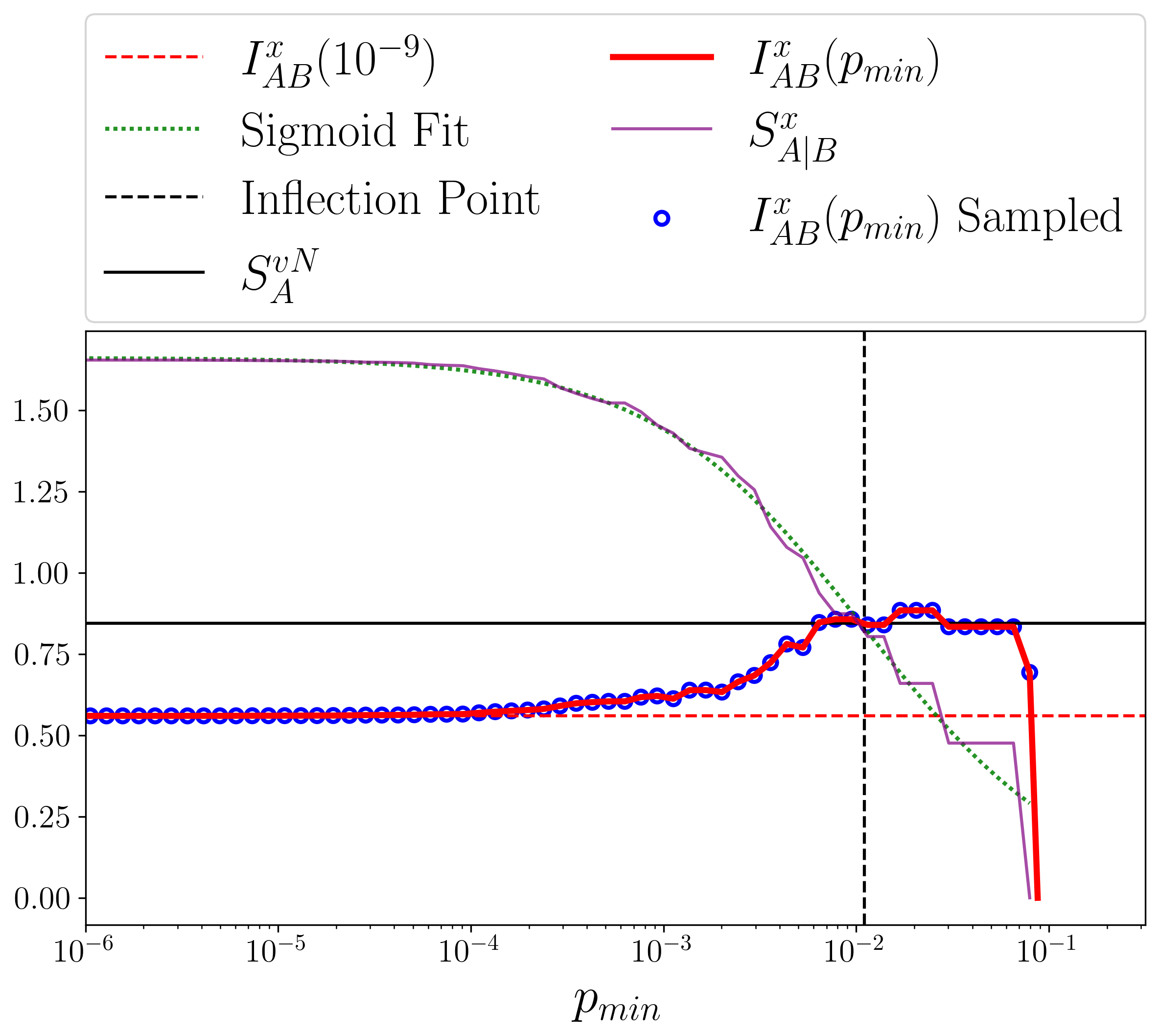} 
\caption{\label{fig:exact6rungsdist} $I_{AB}^X(p_{min})$ for a six-rung ladder with exact diagonalization (continuous line) and DMRG sampling (open circles). Details are provided in the text and in Appendix  \ref{fig1}.}
\end{figure}

In this \art we show that this bound can be improved by ``filtering" the bitstring probabilities, which means removing the measurements which occur less than a given number of times and renormalizing the kept measurements. In the supplementary material of \cite{Meurice:2024fya}, it was observed that when the ramping down of the Rabi frequency before measurement was too slow, large probabilities were amplified and small probabilities reduced. This has resemblance to the filtering described above. It was then shown that for samples of 1000 bitstrings, removing the states with less than 10 observations ($p_{min}=0.01$), increased $I_{AB}^X$ in 6 out of 7 examples considered. In the following, we 
discuss systematically the effect of $p_{min}$ on the mutual information $I_{AB}^X(p_{min})$ as shown in Fig. \ref{fig:exact6rungsdist}.
To our surprise, the conditional entropy $S^X_{A \vert B}$ frequently supplies us with an ideal value for $p_{min}$. $S^X_{A \vert B}$ tends to decrease with filtering along a sigmoid curve, and its inflection point is at a value of $p_{min}$ where $I_{AB}^X(p_{min})$ is much closer to $S_A^{vN}$. A particularly convincing illustration is shown in Fig. \ref{fig:exact6rungsdist}. The horizontal dashed line represents the accurate value of $I^X_{AB}$ without filtering. The purple line is $S^X_{A \vert B}$, and green the sigmoid function that has been fit to it. Both lines interpolate between the upper left and lower right corners.  The inflection point (dashed vertical line) of the sigmoid occurs at $p_{min} \approx 10^{-2}$, where $I_{AB}^X(p_{min})$ has transitioned from the unfiltered  value to a value very close to $S_A^{vN}$ (continuous horizontal line). Fig. \ref{fig:exact6rungsdist} also shows that numerical results obtained with the density matrix renormalization group (DMRG) are very accurate. We will use the DMRG for larger sizes when exact diagonalization is not possible  (see Appendices for more discussion). 

In the following, we introduce the specific model used for calculations, a ladder of Rydberg atoms, and perform calculations with exact diagonalization and DMRG methods \cite{schollwock2005density}, and remote operation of the Aquila facilities \cite{wurtz2023aquila}. We discuss the dependence of the filtered mutual information on the size of the system, the lattice spacing, and the bipartition of the system.
This provides a more nuanced picture than the specific example of Fig. \ref{fig:exact6rungsdist} where the plateaus appear to be very close to the target value.
However, in more general situations, the inflection point method seems to be very efficient to optimize the filtering. This is summarized in the conclusions where we also briefly discuss the implications for experiments with limited resources and discuss to which extent the set of states with low probabilities are featureless.

\section{The simulator: ladder of Rydberg atoms}
Recently, it has become possible to create arrays of Rydberg atoms with adjustable geometries  \cite{Bernien2017Dynamics, Keesling2019Kibble, Labuhn2016RydIsing,Leseleuc2019topo,Ebadi2021_256,Pascal2021AF}.
Their  Hamiltonian reads 
\beq
\label{eq:genryd}
H = \frac{\Omega}{2}\sum_i(\ket{g_i}\bra{r_i} + \ket{r_i}\bra{g_i})-\Delta\sum_i  \hat{n}_i +\sum_{i<j}V_{ij} \hat{n}_i \hat{n}_j,
\enq
with van der Waals interactions $V_{ij}=\Omega R_b^6/r_{ij}^6$,
for a distance $r_{ij}$ between the atoms labelled $i$ and $j$.  We use $\hbar=1$ units. We define the Rydberg occupation $\hat{n}_i\ket{r_i}=\ket{r_i}$ while $\hat{n}_i\ket{g_i}=0$. We follow the convention of using a characteristic length scale, blockade radius $R_b$, at which if two atoms are in the \ket{r_i} state are placed, the interaction strength $V_{ij}$ equals the Rabi frequency $\Omega$. In the following, we focus on a two leg ladder where the rung length $a_y$ is twice the distance between rungs $a_x$, which is also the lattice spacing denoted $a$ hereafter. This choice is motivated by the rich critical behavior observed in Ref. \cite{floating} especially for $R_b/a \simeq 2.35$ for sufficiently large volume, chosen for its proximity to the recently observed floating phase. However, we used open boundary conditions for a rectangular shape rather than the shifted boundary conditions in \cite{floating}. For our use of QuEra's publicly-available Aquila device see  Appendix~\ref{app:aquila}.
\section{Bipartite setup}
%\vskip1pt \noindent 
For a Rydberg array with $N_q$ atoms either in the ground $\ket{g}$ or Rydberg $\ket{r}$ state, the whole system ($AB$) can be divided into two subsystems $A$ and $B$. The computational basis consists of the $2^{N_q}$ elements 

\beq
\ket{\{n\}}\equiv\ket{n_0, n_1,\dots,n_{N_q-1}},
\enq
with $n_j=$ 0 or 1 representing $\ket{g}$ and $\ket{r}$ respectively. 
Note that some authors use the opposite convention. Any element of this basis can be factored in a bipartite way by splitting the $N_q$ qubits into $|A|$ and $|B|$ qubits with $|A|+|B|=N_q$. 
\beq
\ket{\{n\}_{AB}}=\ket{\{n\}_A}\ket{\{n\}_B},
\enq
with
\begin{eqnarray}
\ket{\{n\}_A}&\equiv&\ket{n_0, n_1,\dots,n_{|A|-1}} \ {\rm and } \\
    \ket{\{n\}_B}&\equiv&\ket{n_{|A|},\dots,n_{N_q-1}}
\end{eqnarray}
Given an arbitrary prepared state $\ket{\psi}$,
we can write it as a superposition of the states in the computational basis
\beq
\ket{\psi}=\sum_{\{n\}} c_{\{n\}}\ket{\{n\}}. 
\enq

%Writing the $2^{N_q}$ dimensional vector $c_{\{n\}}$ corresponding to the vacuum as a $2^{|A|}\times 2^{|B|}$
%matrix 
%\beq
%C_{\{n\}_A ,\{n\}_B} =c_{\{n\}}, 
%\enq
%we find that 
Using $\rho_{AB}=\ket{\psi}\bra{\psi}$, the reduced density matrix of the subsystem $A$ is defined as $\rho_A=\mathrm{Tr}_B( \rho_{AB})$ and 
%can be written as 
%\beq
%\rho_{A \{n\}_A ,\{n'\}_A} = (C C^{\dagger})_{\{n\}_A ,\{n'\}_A},
%\enq
%in the computational basis. 
the corresponding von Neumann entanglement entropy
\beq
S_A^{vN}=-\mathrm{Tr}\,(\rho_A \ln (\rho_A))=-\sum_m \lambda_m \ln(\lambda _m), 
\enq
where $\lambda_m$'s are the eigenvalues of $\rho_A $ which are independent of the basis used in $A$. In an experiment, 
the state $\ket{\{n\}}$ is observed with a probability
\beq
p_{\{n\}}=|c_{\{n\}}|^2.
\enq
We define the associated Shannon entropy as 
\beq
S_{AB}^{ X}\equiv -\sum_{\{n\}} p_{\{n\}}\ln(p_{\{n\}}).
\enq
Following Shannon \cite{shannon1949mathematical} we define the marginal probability in the subsystem $A$ by summing over $B$:
\beq
p_{\{n\}_A}=\sum_{\{n\}_B}p_{\{n\}_A \{n\}_B},
\enq
and the corresponding entropy
\beq
S_{A}^{X}\equiv -\sum_{\{n\}_A} p_{\{n\}_A}\ln(p_{\{n\}_A}).
\enq
Similarly, we can define $S_{B}^{X}$ by interchanging $A$ and $B$ in the above equations. 
The classical mutual information is defined as
\beq
\label{eq:mi}
I^X_{AB}\equiv S_A^X+S_B^X-S_{AB}^X.
\enq
It was found by Shannon \cite{shannon1949mathematical} that this quantity is non-negative. Upper bounds can also be found by reducing $\rho_A$ to the diagonal part $\rho^X_{A}$ \cite{jpch10,wittenmini}. 
In summary \cite{mehdi,Meurice:2024fya,Yang:2024yxu} we have the bounds:
\beq 0\leq I^X_{AB}\leq S_{B}^{vN}=S_{A}^{vN}\leq S_A^X ({\rm or} S_B^X).
\label{eq:bound2}
\enq
Illustrations of the behavior of these quantities as functions of the adjustable parameters are given in Appendix~\ref{app:ent} for the case of a six-rung ladder. The numerical methods used to calculate the vacuum state and sample the probabilities are presented in Appendix~\ref{app:num}. The errors associated with the 
finite size of the sampling are illustrated in Appendix~\ref{app:flucdmrg}.

%\vfill
%\eject

\section{Where should we stop? }%\vskip1pt \noindent 
Since the filtered mutual information is not always a lower bound of the unfiltered $S_A^{vN}$, we need to determine the optimal value of $p_{min}$.  Fig. \ref{fig:exact6rungsdist} suggests that we could stop when $I_{AB}^X$ stops increasing over a ``significant" range, but the other included plots show that this is not generally a reliable indicator. 
Note that there are short range fluctuations due to the discrete nature of the probabilities which are clearly seen in Appendix~\ref{app:aquila}. 
%Efforts to automatize this process are under %investigation.  Current efforts center around using %the second derivative of $I^X_{AB}$ vs. $p_{min}$ %graphs to identify areas of relative plateau as %candidate locations for improving on an unfiltered %bound.   

\begin{figure}
\includegraphics[width=7cm]{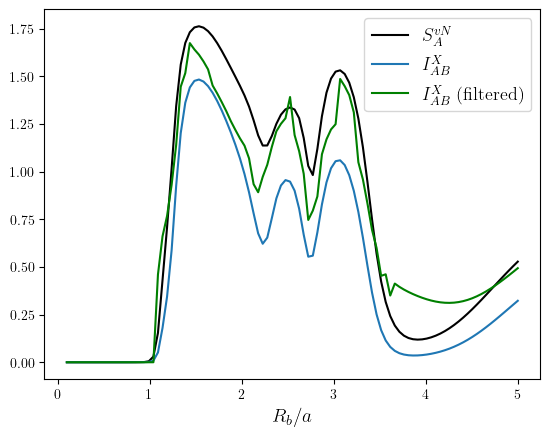} 
\caption{\label{cond_ent_comp} Comparison of filtered (green) and unfiltered (blue) mutual information as estimates of entanglement entropy (black) for 6 rungs, $\Delta/\Omega=3.5$.}
\end{figure}

We have observed the conditional entropy $S^X_{A \vert B}$ between the subsystems to provide another indication of when to stop filtering. It is also calculated from the bitstring data and is the difference between the upper and lower bounds in Eq. (14).
\beq S^X_{A\vert B} \equiv S^X_{AB} - S^X_B
\enq
The conditional entropy is another measure of correlation between two random variables and is a part of calculation of mutual information.
\beq I^X_{AB} = S_A - S^X_{A\vert B}
\enq
Its value decreases under filtering as the number of states in the distribution diminishes. In many cases, this curve can be well fit with a sigmoid and contains an inflection point, where the change in conditional entropy begins slowing down. We hypothesize that this inflection point is a good place to stop the filtering, and demonstrate its utility here with several examples. Fig. \ref{cond_ent_comp} shows the improved estimate across several phases of the ladder. This position also corresponds roughly to where the conditional entropy is reduced by half from its original value. This filtering method tends to work better in higher entropy regions of the phase space, and especially well near phase transitions. The overestimation of $S^{vN}$ at large $R_b/a$ most likely stems from the amplified significance of the edges. For larger volume, we expect a better fit in that regime. Note that, unlike $S^{vN}$ and $I^X_{AB}$, $S^X_{A \vert B}$ is not equivalent between A and B when the bipartition is not equal size (visible in Fig. \ref{fig:VaryingSystemSizeopt}). Future work will further investigate these estimation improvements \cite{Senseman Future}. 

In many cases, the filtered $I^X_{AB}$ eventually rises above $S^{vN}_A$. This does not violate the Holevo bound because, as explained in Appendix~\ref{fig1}, filtration of the quantum state, performed with a projection operator, affects $S^{vN}_A$.
\section{Volume dependence}
\begin{figure}[h]
\includegraphics[width=8.cm]{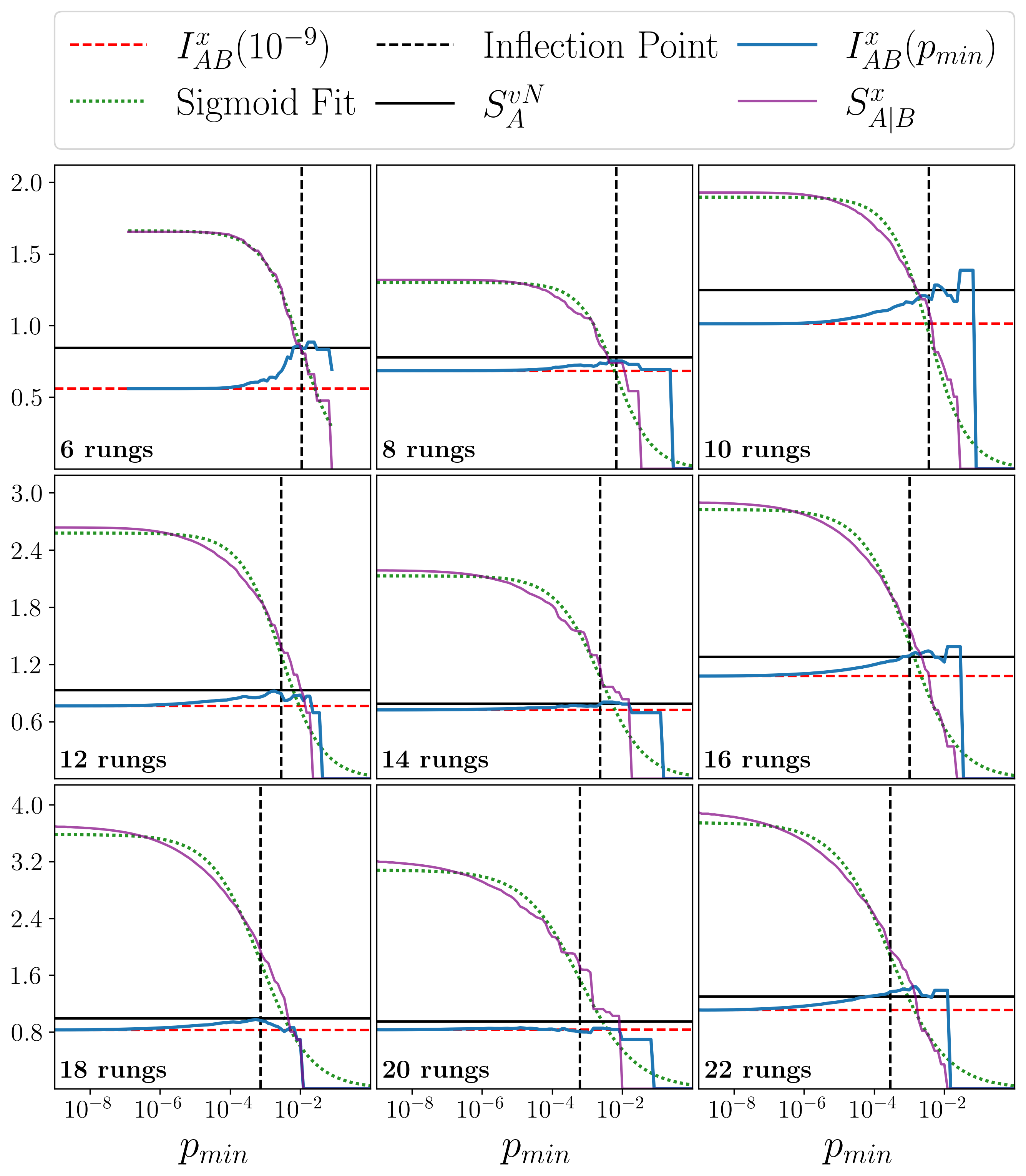} 
\caption{\label{fig:VaryingSystemSize} Effects of the system size on $I_{AB}^X(p_{min})$ for 6, 8, ..., 22 rungs with $R_b/a=2.35$ and 1 billion counts obtained via DMRG.}
\end{figure}

%\vskip1pt \noindent 
%\vskip2pt \noindent
We repeated the procedure described for Fig. \ref{fig:exact6rungsdist} with more rungs.
The results are shown in Fig. \ref{fig:VaryingSystemSize}. 
Up to 18 rungs, filtering the mutual information up to the inflection point of the conditional entropy obtains better estimates of $S_A^{vN}$ than the original. For 20 rungs, $I_{AB}^X(p_{min})$ decreases over a large part of the left side and this suggests that the best available estimate is the unfiltered value, which appears to be a rather tight lower bound. For 22 rungs, the proposed filtering slightly overshoots the exact $S_{A}^{vN}$ but still reduces the error compared with the unfiltered lower bound.
%\vfill
%\eject

%\vfill
%\eject
\section{Lattice spacing dependence}
We have considered the effect of changing the lattice spacing on $I_{AB}^X(p_{min})$. For $R_b/a =1.00$, the entanglement is very low and so is $I_{AB}^X(p_{min})$ over a broad range on the left. In cases like these, there is no benefit to filtering, but that is manifest in the original extremely low value for correlations. For $R_b/a$ between 1.25 and 2.25, there are some significant plateaus at values between the unfiltered $I_{AB}^X$ and $S_A^{vN}$. For $R_b/a$ between 2.5 and 3.00, there are also broader plateaus at values very slightly above $S_A^{vN}$. In most cases, our proposed stopping point based on $S^X_{A \vert B}$ picks out a good choice.
\begin{figure}[h]
\includegraphics[width=8.6cm]{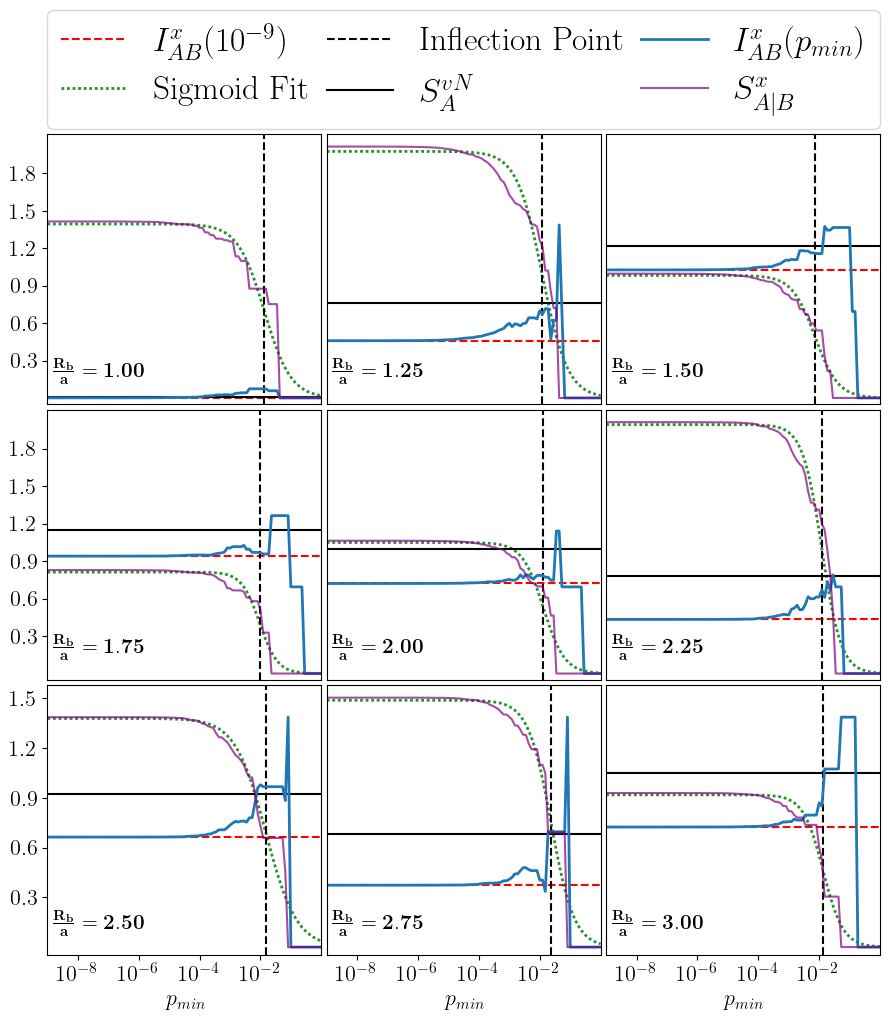} 
\caption{\label{fig:VaryingRba}
Effects of the lattice spacing  on $I_{AB}^X(p_{min})$ for $R_b/a=1.0, 1.25, \dots, 3.0$ on 6 rungs obtained via Exact Diagonalization.}
\end{figure}

\section{Bipartition dependence. }
\begin{figure}[h]
\includegraphics[width=8.6cm]{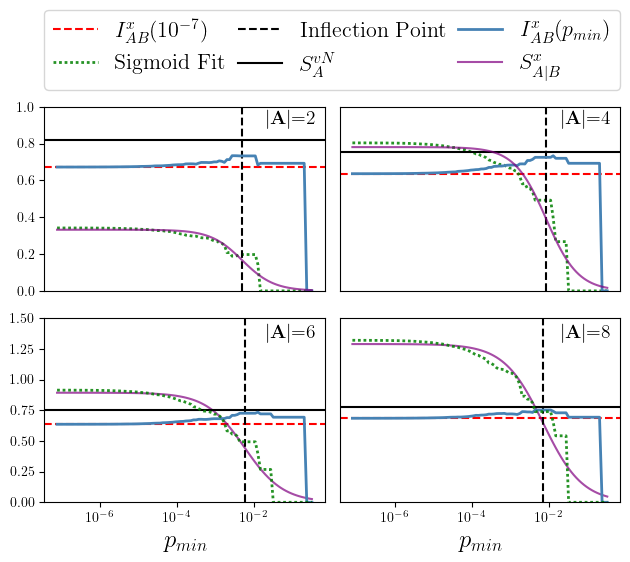} 
\caption{\label{bipartition_plots}Illustration of the effects of filtering for differing subsystem sizes for a bipartition.  This behavior mirrors for equivalent sizing.  Probabilities obtained with 10 million counts of DMRG.  Plot made for an eight-rung system with $R_b/a=2.35$.}
\end{figure}

%\vskip2pt \noindent
%ZANE
This method of estimating entanglement entropy is not limited to just bipartitions with $|A|=|B|$.  This is shown in Fig. \ref{bipartition_plots} where a Rydberg ladder of 8 rungs and $R_b/a=2.35$ is investigated.  The methodology is just the same as before, using Eq. (\ref{eq:mi}), but now the sizes $|A|$ and $|B|$ need not be equal.  The entanglement entropy and the mutual information are both symmetric under the interchange of $A$ and $B$ and subsystems with $|A|$ = 2 and $|B|$ =14, 4 and 12, etc. provide identical results.   When incorporating the filtered probabilities with varying subsystems it is observed that the filtration reaches closer to the von Neumann entropy when the von Neumann entropy is lower. This behavior is shown in Fig. \ref{bipartition_plots} with the inflection point of the conditional entropy fit for all subsystem sizes at this volume.   %in Fig. \ref{bipartition_plots}.  To find this %p_{min}$ exactly in a routine way will be a goal of %future work.

\section{Analog simulations with Aquila}
\begin{figure}[!htb]
        \centering
        \includegraphics[width=0.45\textwidth]{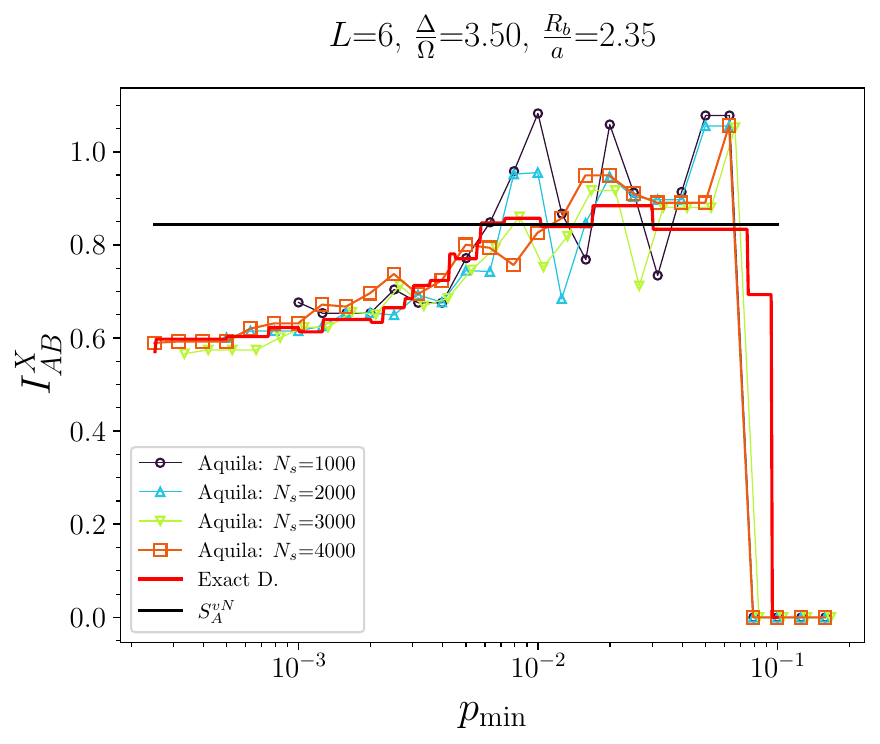}
        \caption{The number of shots is varied to investigate the deviation of the Aquila simulation result with the exact diagonalization.}
    \label{fig_shot_Aquila}
    
\end{figure}

%\vskip2pt \noindent
So far we relied on numerical methods to obtain vacuum states and their associated probabilities.
In this subsection, we discuss results obtained by adiabatically preparing the vacuum states and then performing measurements using Quera's Rydberg atom device, Aquila.
See Appendix~\ref{app:aquila} for more information. 
As a first step, we considered the case of 6 rungs with $R_b/a=2.35$ and compared with the exact results of Fig. \ref{fig:exact6rungsdist}.
We performed a comparison of the variation in shot counts for the computation of filtered mutual information. It is clear from the comparison with the exact diagonalization results in Fig.~\ref{fig_shot_Aquila}, that increasing the number of shots from 1000 to 4000  tends to improve the filtered mutual information $I^X_{AB} (p_{\mathrm{min}})$. However, aside from the statistical error, there are several sources of errors including waveform error, detection error, and scattering from the intermediate state due to lasers \cite{Ebadi2021_256}.  In some cases the discrepancies with numerical results are significant and cannot be understood as statistical errors. 
The sources of these errors and potential mitigation strategies are currently under investigation. More results obtained from Aquila, with other lattice spacings and number of rungs are provided in Appendix~\ref{otheraquila}.

\section{Estimates of $S_A^{vN}$ with a few thousands shots} 
In our numerical calculations, we were able to have either all the probabilities using exact diagonalization, or millions of shots distributed in good approximation as in the exact results when using the DMRG. However, as far as improving the mutual information bound, the states with low probabilities do not play an important role. Typically, we observe significant onsets near $p_{min}\sim 10^{-4}$ which suggests that we can get improved estimations of $S_A^{vN}$ using only a few thousand shots. This is illustrated in the example of Fig. \ref{fig_shot_Aquila}, where we approximately reproduce the exact filtered mutual information $I_{AB}(p_{min})$ with $p_{min}\geq 10^{-3}$.
This is good news, because when using actual physical quantum devices, one is often limited to extracting information from a few thousand shots.

\begin{figure}[h]
\includegraphics[width=8.9cm]{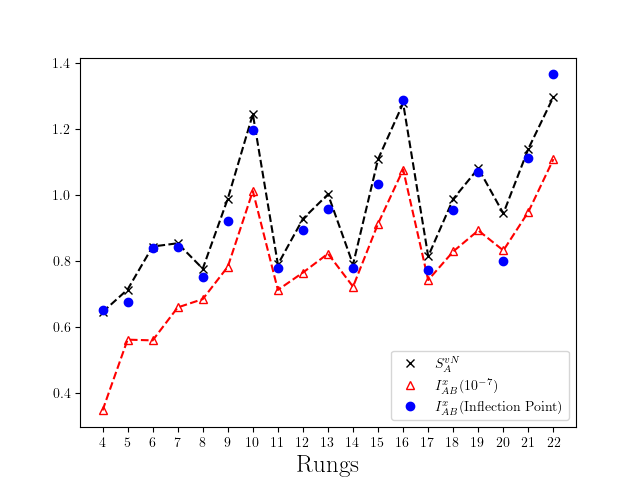} 
\includegraphics[width=8.cm]{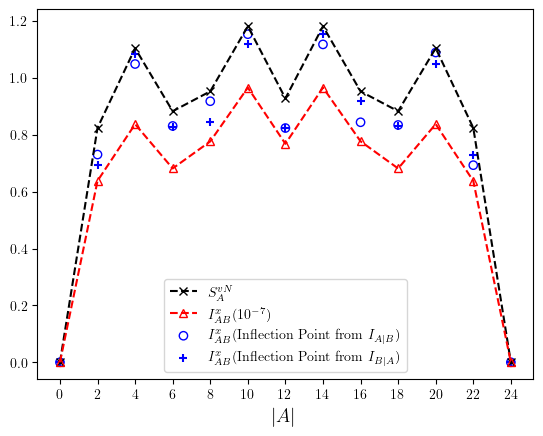} 
\caption{\label{fig:VaryingSystemSizeopt} $I_{AB}(10^{-7})$ (triangles), $I_{AB}$(Inflection Point) (circles and/or pluses) using the sigmoid fit inflection point as the best filtering spot, and $S_A^{vN}$ (crosses) for $R_b/a=2.35$ and 10 million shots obtained via DMRG. Top: for 4, 5, ...., 22 rungs with even bipartitions; Bottom: for 12 rungs with different bipartitions where $|A|$ is the number of atoms in one partition.}
\end{figure}

%update below here
As additional illustrations, we have collected $I_{AB}(p_{min})$ at the inflection point for $R_b/a=2.35$ with different numbers of rungs with equal bipartitions 
and 12 rungs for various bipartitions (Fig. \ref{fig:VaryingSystemSizeopt}).  First, this figure highlights how the conditional entropy given the A or the B region differs.  Both will provide an improved lower bound, but they are symmetric with region size providing different improvement given the conditional entropy of which region is used.  In all but three cases, this simple procedure provides values above the unfiltered mutual information and below the target value. These two figures rely on DMRG calculations with $10^7$ shots. While these filter values provide an improvement on the mutual information, the gain comes with simulations with a larger number of shots.  With a smaller number of shots certain states will occur less frequently, increasing the error on estimated probabilities. Fig. \ref{fig:VaryingSystemSizeopt} shows that for the examples considered here, choosing a $p_{min}$ based on the inflection point of the conditional entropy fit tends to provide significant, but not always optimal, improvements allowing estimation of the quantum entanglement with limited resources. 

\section{Conclusions}
Using examples of ladders of Rydberg atoms, we have shown that by excluding bitstrings with probabilities lower than $p_{min}$ and renormalizing the remaining data, a procedure that we call filtering, we tend to obtain larger mutual information and better approximations to the von Neumann entanglement entropy $S_A^{vN}$. 
%Current efforts are underway to automatize the selection of %an improved filtration level to achieve a filtered level %closer to the true entanglement entropy. 

We demonstrate the conditional entropy to provide one very effective filtration paradigm. In a significant number of cases considered, filtering to the value of $p_{min}$ corresponding to the inflection point in the conditional entropy yields a much tighter approximation to the entanglement entropy than stopping at plateaus. This method shows remarkable improvement across a large range of volumes, lattice spacings, and partitions. 
%works in a large majority of cases expect when I decreases %which often corresponds to a tight upper bound?
%The reasons for this behavior are not well understood, but %some intuitive support can be given.

We may view the low-probability states as featureless as far as the diagonal part of the density matrix is concerned, and that removing them provides a better estimate of the quantum entanglement. The reduction of the conditional entropy can serve as a quantifier of this cleaning up of the distribution, and stopping midway preserves much of the information about the original state.

We are currently trying to make this picture 
more quantitative by comparing averages of observables for sets of states in specific probability regions. Using observables such as the double Rydberg occupation on a rung we can see that  
%The picture that emerges is the following. 
%In other words, it is 
the set of states with very low probabilities tends to be featureless. By this, we mean that the averages of observables are very different from the vacuum average and standard deviations are large. As we filtrate, important features of the state become more apparent. If we proceed further and retain only a few states with the largest probabilities, we obtain an oversimplified picture which ignores fluctuations. This description is supported by the calculation of the conditional entropy. The featureless behavior is characterized by large values and the oversimplified behavior by low values of the conditional probability. 
There is a transition region where the mutual information appears to provide very good estimates of the quantum von Neumann entropy.

Analog simulations with a few thousands shots provide a limited access to states with low probabilities and naturally provide 
significantly filtered data. The origin of the discrepancies between Aquila data and numerical methods is currently under investigation. More specifically, we are characterizing errors due to imperfect state preparation, $\Omega$ ramping  down, readout errors  and statistical errors. 
Future work may extend these techniques to larger systems and more complex geometries, furthering their applications for analog simulations.

\section*{Acknowledgements} 
%\vskip2pt \noindent
This research was supported in part by the Dept. of Energy
under Award Number DE-SC0019139 and DE-SC0025430. We thank 
the Amazon Web Services and S. Hassinger for facilitating remote
access to QuEra through the Amazon Braket while teaching quantum mechanics and our Department  of Physics and Astronomy for supporting part of the cost of the analog simulations presented here. We thank the University of Iowa for providing access to the Argon computing facilities. We acknowledge the use of NERSC facilities with 
ERCAP 0023235 award. Discussions during the long-term workshop, HHIQCD2024, at Yukawa Institute for Theoretical Physics (YITP-T-24-02), were useful as we completed this work.
The final version was completed while Y. M. attended the workshop ``Co-design for Fundamental Physics in the Fault-Tolerant Era" at the 
InQubator for Quantum Simulation, University of Washington, Seattle, USA and he benefited from discussions with the participants. M.A. would like to thank S. Hassinger, T. Aunon, N. Owens and others on the on-boarding team for arranging quantum credits for the Aquila device through amazon braket.

\section*{Data Availability}

The datasets generated and analyzed during the current study are available in the Zenodo repository~\cite{zenodo}. Numerical data was prepared by Corona~\cite{corona_data}.  
The source code used for data processing and analysis is available in the QCOM repository~\cite{kaufman_qcom_v020}.

\appendix
\section{Detailed features of Fig. 1} 
\label{fig1}In this Appendix, we discuss specific aspects of Fig. 1 in the main text.
We calculated 
the vacuum of a six-rung ladder with $R_b/a$=2.35. 
This first choice of lattice spacing was motivated by the rich critical behavior found in Ref. \cite{floating}. 
There are two atoms on each rungs of the ladder or equivalently six atoms on both legs, and so we have twelve atoms for the the entire system. 
We first used exact diagonalization and extracted an accurate value for $S_A^{vN}\simeq 0.844$ (top horizontal line in Fig.\ref{fig:exact6rungsdist}) for $A$
being the 3 rungs on say the left side of the ladder. 
We then calculated the probabilities to observe any of the 4096 states available in the computational basis and obtained the mutual information $I_{AB}^X\simeq 0.559$ which is about 30 percent below $S_A^{vN}$ (dashed line). We then removed the states with a probability less than $p_{min}$, starting with $p_{min}=10^{-6}$, renormalized the remaining probabilities and then recalculated the filtered $I_{AB}^X (p_{min})$. 

We then repeated the vacuum calculation using MPS and the DMRG method to generate $10^7$ random states with probabilities corresponding to the $|c_n|^2$ using a recursive method available with ITensor (see Section \ref{app:num}). 
This can be thought to be an ideal experiment with $10^7$ shots.
The results are in excellent agreement with the exact diagonalization which validates the DMRG method that will be used for larger volumes when exact diagonalization becomes difficult.

\begin{figure}[h]
\includegraphics[width=8.6cm]{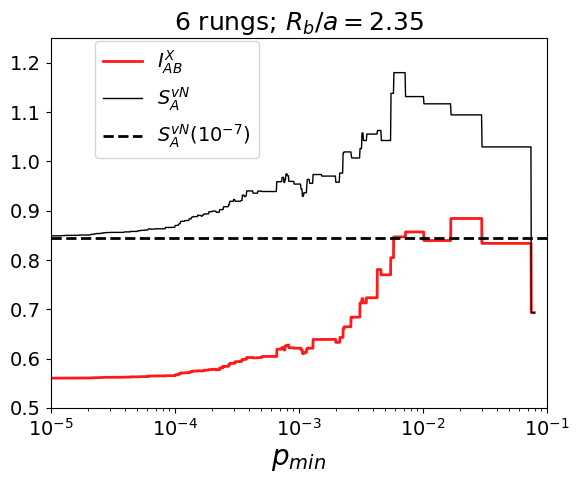} 
\caption{\label{fig:trunc6rungsdistE} 
 $I_{AB}^X(p_{min})$ and $S_{A}^{vN}(p_{min})$ for a six-rung ladder with exact diagonalization.}
\end{figure}

Attentive readers may be concerned that the tendency for the filtered $I^X_{AB}$ to exceed the $S^{vN}_A$ in some regions of $p_{min}$ is a violation of the Holevo bound. However, the filtering corresponds to the removal of states, which will also affect $S^{vN}_A$. To see this, we introduce a projector corresponding to a given filtration
\beq
P(p_{min})=\sum_{\{ n\} : p_{\{ n\}}\geq p_{min} }\ket{\{ n\}}\bra{\{ n\}}.
\enq
Given a state $\ket{\psi}$, we define the normalized projected state as 
\beq
\ket{\psi, p_{min}}=P(p_{min})\ket{\psi}/\sqrt{\bra{\psi}P(p_{min})\ket{\psi}}.
\enq
Starting from the original vacuum, we can define the filtered density matrix 
\beq\rho_{AB}(p_{min})=\ket{vac.,p_{min}}\bra{{vac.,p_{min}}},\enq
its reduced version
\beq\rho_A(p_{min})=\mathrm{Tr}_B( \rho_{AB}(p_{min}))),\enq
and we then calculate
\beq
S_A^{vN}(p_{min})=-\mathrm{Tr}\,(\rho_A (p_{min})\ln (\rho_A(p_{min}))).
\enq
We have the bound
\beq I^X_{AB} (p_{min})\leq S_{B}^{vN}(p_{min})=S_{A}^{vN}(p_{min}).
\label{eq:boundf}
\enq
which is clearly obeyed in Fig. \ref {fig:trunc6rungsdistE} and there is no contradiction. Moreover, the distance between the two filtered quantities stays more or less constant ($\simeq 0.3$). We are interested in estimating the full $S^{vN}_A$ without filtration and the actual value of 
 $S^{vN}_A(p_{min})$ does not seem relevant for this purpose.

\section{Quantum device simulation methods}
\label{app:aquila}
We conducted remote analog quantum simulations using QuEra's Aquila device \cite{wurtz2023aquila}. This reference contains the dominant sources of error; laser noise, atom motion, state decoherence, inhomogeneity, and issue in measurements. The motivation behind this is to investigate larger system sizes which at some point will become inaccessible with classical computation methods. With the naive exact diagonalization we could investigate a Rydberg system of size $N_q=12$ atoms, whereas with high-precision DMRG calculation we could investigate up to a system size $N_q=44$ atoms. In the near future, we will be able to investigate models with far more atoms than this with the quantum device. Aquila has a roadmap for extending their device to simulate more than 3000 atoms in the near future \cite{quera_roadmap}. 
\begin{figure}[h]
\includegraphics[width=8.6cm]{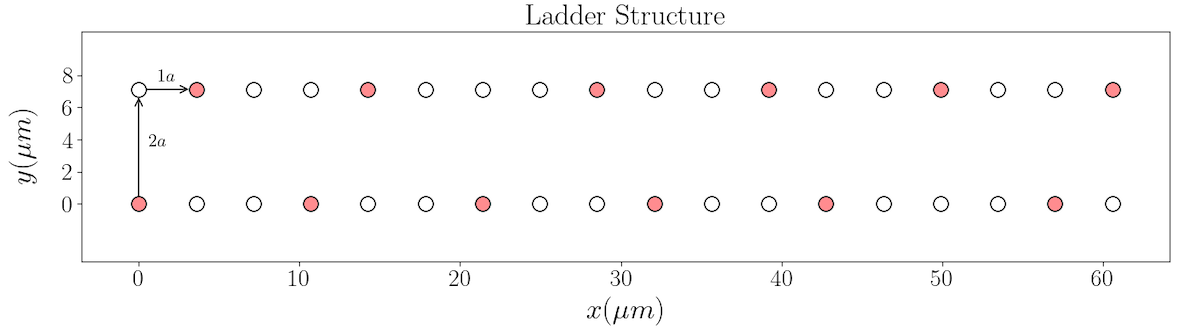} 
\caption{\label{fig:ladder}Illustration of the lattice structure for an eighteen-rung ladder with ground state (empty circle) and Rydberg state (red circle) atoms.}
\end{figure}

We prepared the ground state of the system adiabatically. Initially, we prepare a state with all the atoms in the ground state. Incidentally, this is the ground state of the Rydberg system at infinite negative detuning and zero Rabi frequency. Next, we slowly tune the Rabi frequency from zero to the target frequency ($\Omega$) and in the same duration, we increase the detuning from a large negative value to the target value ($\Delta$). Ideally, due to the slow adiabatic evolution we can reach the ground state of the target parameters as the system is gapped. Before any measurement is performed, Rabi frequency is switched off as this is essential for the fluorescence imaging in the readout process \cite{Bernien2017Dynamics}. This is done as fast as possible so that the ground state of the target Hamiltonian is minimally disrupted by other low-lying energy states.  There are different choices on how to tune the parameters in the Rydberg simulation to prepare ground state at different phases. Once we set the lattice spacing $a$ to a fixed number, we can set the final Rabi frequency and final detuning to different values to recover the ground state at different phases.
For the Aquila simulation, we use the same parameters to that of DMRG simulation in all the cases, except for the case $R_b/a=3.0$, where we use $\Omega=0.4982\pi$ MHz due to the geometrical constraint that distance between two nearest-neighbor atoms needs to be more than 4 $\mu$m at the current publicly available setup at the Aquila device.

\section{Phase diagram and bound on entanglement entropy\label{app:ent}}
\begin{figure}[h]
\includegraphics[width=0.49\textwidth]{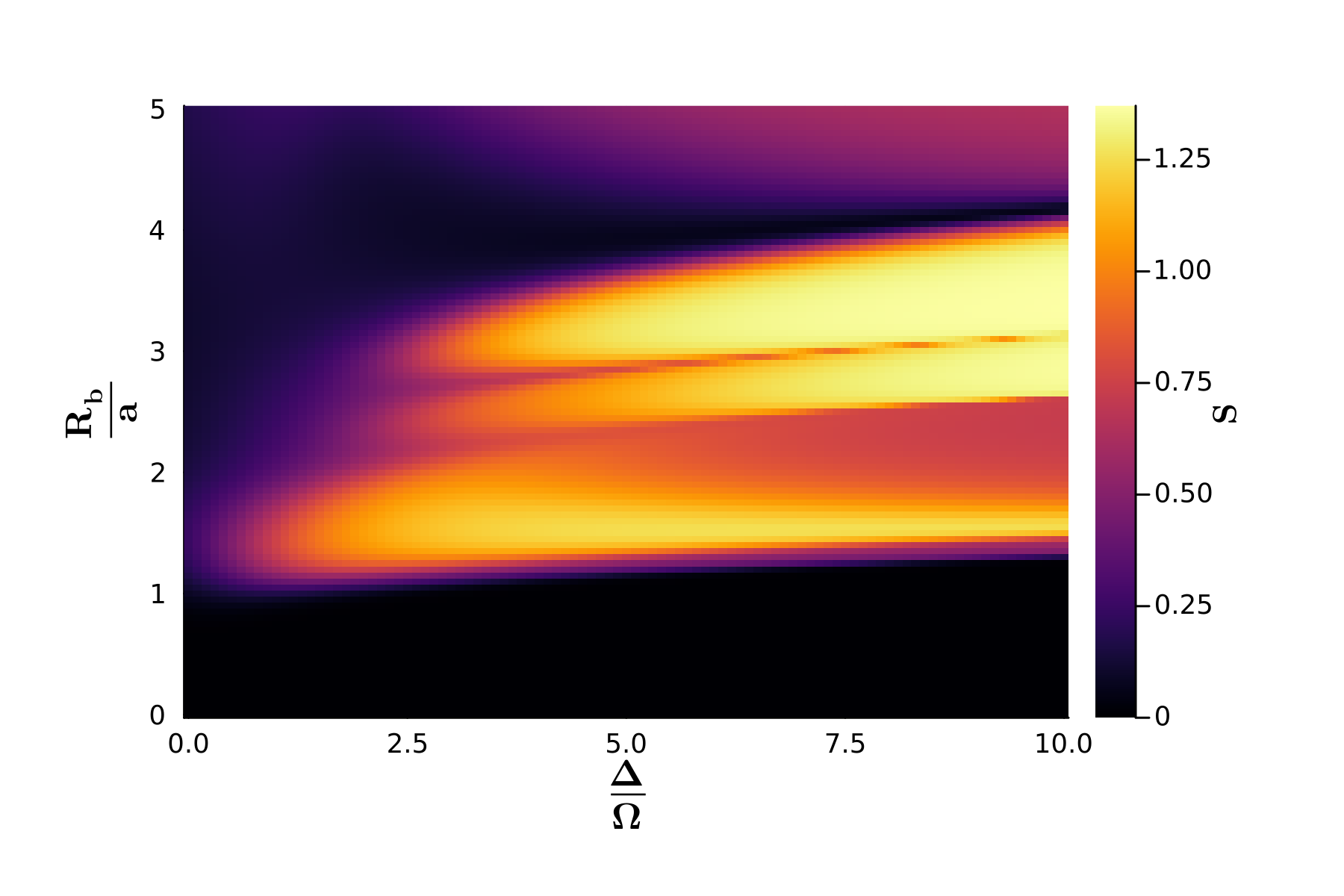} 
\caption{\label{fig:phases} von Neumann entanglement entropy for a six-rung ladder with $a_y=2a_x$ as a function of $R_b/a_x$ and $\Delta/\Omega$.}
\end{figure}

In this, we present the entanglement entropy for a six-rung ladder as a function of $R_b/a$ and $\Delta/\Omega$. 
The results are shown in Fig.~\ref{fig:phases}. For small systems, the results
are sensitive to the system size and boundary conditions. 
For a detailed overview on different phases in the Rydberg ladder at larger volumes and different boundary conditions see \cite{floating}.

\begin{figure}[h]
\includegraphics[width=0.45\textwidth]{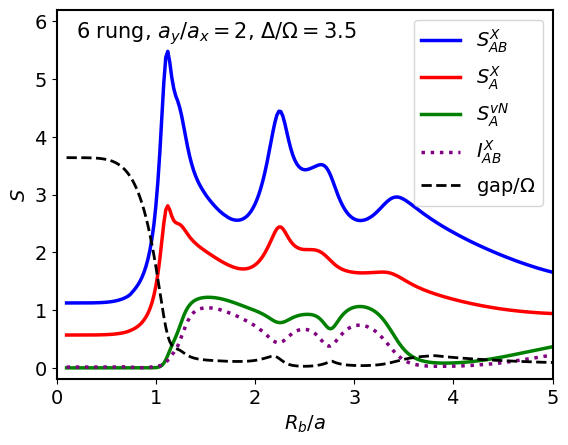} 

\caption{\label{fig:holevo} 
Comparison of $S_{AB}^X$, $S_A^X$, $S^{vN}_A$ and $I^X_{AB}$ at a fixed  $\Delta/\Omega=3.5$.}
\end{figure}

In Fig.~\ref{fig:holevo}, we plotted the 
bitstring entropy $S^X_{AB}$, the 
reduced bitstring entropy $S^X_A$, the von Neumann entropy $S^{vN}_A$ and the classical mutual information $I^X_{AB}$ as we vary the inverse lattice spacing in blockade-radius unit. We observe agreement with the rigorous bounds 
$ 0\leq I^X_{AB}\leq S_{A}^{vN}\leq S_A^X$. We also observe that $S^{vN}_A$ 
becomes larger when the gap closes.

\section{Numerical methods}
\label{app:num}
Numerical calculations were performed with 
$\Omega=5\pi$ MHz, which implies $R_b=8.375 \mu$m, and a detuning $\Delta=17.5\pi$ MHz as in \cite{floating}. $R_b/a$ was varied between 1 and 3. Exact diagonalization calculations were performed in python using the linalg library.  
The DMRG calculations  of vacua and the samples of measurements with the corresponding probability were performed with the ITensor software  library in Julia \cite{itensor}. 

\section{Fluctuations in finite sampling}
\label{app:flucdmrg}

\begin{figure}[h]
    \centering
    \includegraphics[width=1\linewidth]{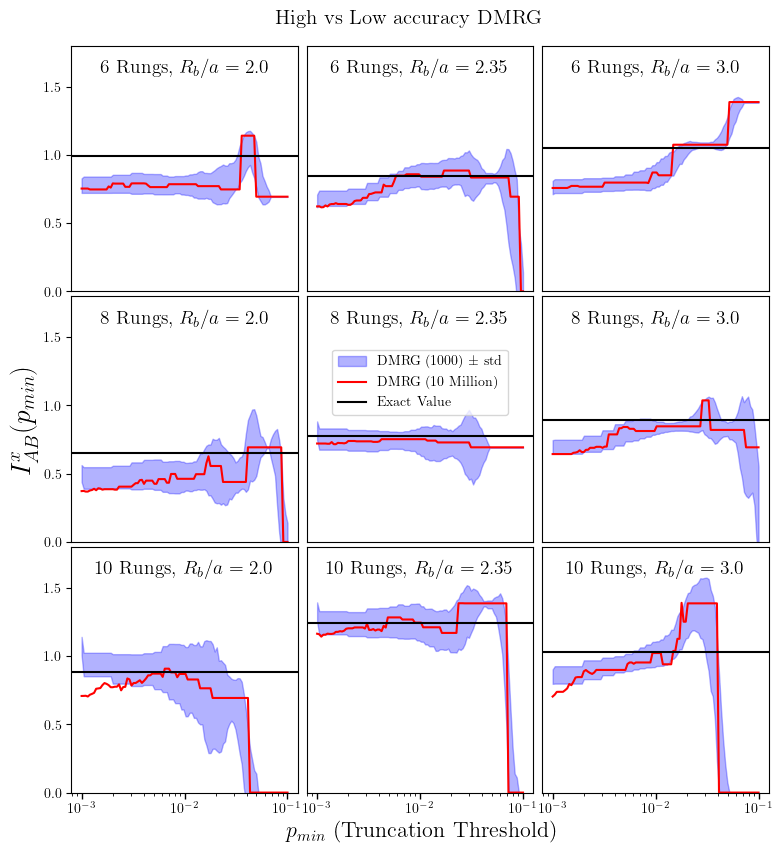}
    \caption{\label{fig_DMRG}DMRG (10 million counts) vs DMRG (1000 shots with error) for various $R_{b}/a$ and system sizes, compared to the numerically  accurate value of the von Neumann entropy computed from the DMRG calculation. }
\end{figure}

With the current technology, the number of shots obtained in analog simulations is much lower than the 10 million shots obtained with the DMRG. Using QuEra's Aquila device \cite{wurtz2023aquila} we are able to produce 1000 shots on a consistent basis. In order to produce DMRG data that resembles experimental output, we randomly reduce our sample size to 1000 shots. To see the impact a smaller sample size has on the error of the mutual information, we do this smaller sample size 1000 times. From these 1000 samples of 1000 shots we calculate error bars. Fig. \ref{fig_DMRG} shows the high accuracy DMRG (10 million shots) plotted against the Sampled DMRG (1000 shots) with error bars. The figure shows that the randomly selected DMRG subsamples roughly follow the shape of the higher accuracy DMRG. This puts the analog calculation discrepancies in perspective. \\

\section{Other analog simulations}\label{otheraquila}

\begin{figure}[h]
    \centering
    \includegraphics[width=1\linewidth]{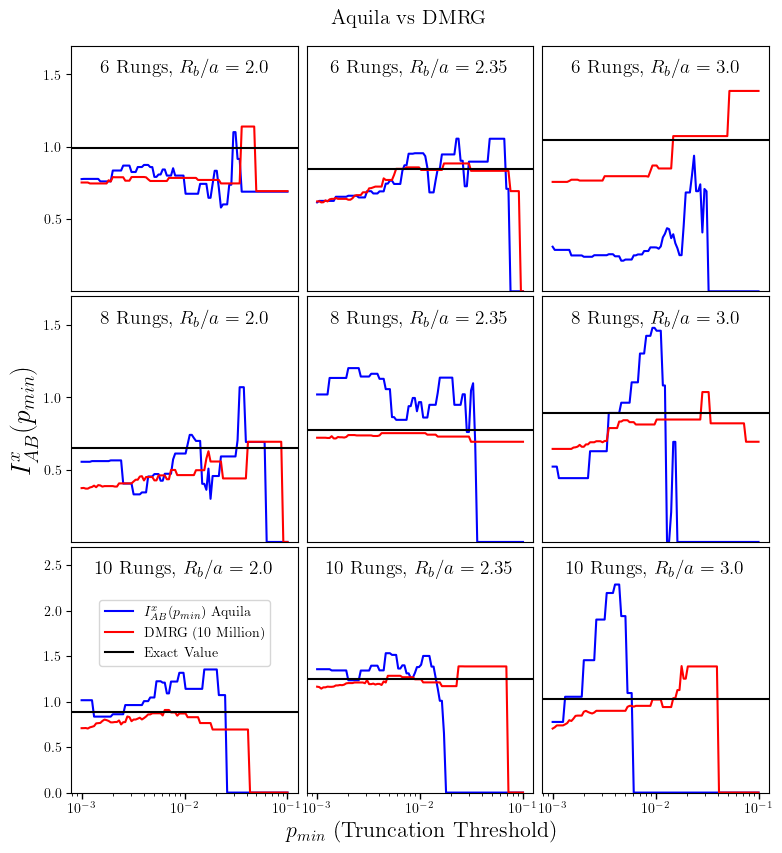}
    \caption{ Filtered mutual information obtained from DMRG ($10^7$ samples of ground states) vs Aquila (2000 shots) for various  $R_{b}/a$ (2.0, 2.35 and 3.0) and system sizes (6, 8 and 10 rungs), compared to the numerically accurate  value of the von Neumann entropy.\label{fig_Aquila}}
\end{figure}

We performed analog quantum simulations with Aquila for three different sizes of the ladder and at three different $R_b/a$ at a fixed $\Delta/\Omega=3.5$. 
The results are shown in  Fig.~\ref{fig_Aquila}. We observe that some of the qualitative features of the filtered classical mutual information are consistent with the DMRG computation.
In some cases, we observe an increase in the filtered mutual information as we increase $p_{min}$. However, the position of the peak in the filtered entropy doesn't always coincide with the DMRG computation. The discrepancy is smaller for the shorter ladders. 

Since the process of filtering probabilities involves characterizing all states including low-probability states, and the Hilbert space grows exponentially with the number of atoms ($\mathrm{dim}(H) \propto 2^{N_q}$), recovering quantitative results for the filtered mutual information becomes challenging with low statistics. A quantitative analysis of the statistical requirements on the volume warrants further study. We also note a larger discrepancy as the ratio of $\frac{R_b}{a}$ is increased, this is predominantly due to the larger errors associated in the laser waveforms for cases with higher ratio. A detailed study of this issue is currently being investigated and authors will present the results elsewhere.

\clearpage

\end{document}